\documentstyle[12pt]{article}

\begin{document}

\begin{titlepage}
\null\vspace{-62pt}

\pagestyle{empty}
\begin{center}

\vspace{1.0truein} {\Large\bf The path integral measure, constraints and ghosts for massive gravitons with a cosmological constant}

\vspace{1in}
{\large\bf Dimitrios Metaxas} \\
\vskip .4in
{\it Department of Physics,\\
National Technical University of Athens,\\
Zografou Campus, 15780 Athens, Greece\\
metaxas@central.ntua.gr}\\

\vspace{0.5in}

\vspace{.5in}
\centerline{\bf Abstract}

\baselineskip 18pt
\end{center}

\noindent
For massive gravity in a de~Sitter background one encounters problems of stability when the curvature is larger than the graviton mass. I analyze this situation from the path integral point of view and show that it is related to the conformal factor problem of Euclidean quantum (massless) gravity. When a constraint for massive gravity is incorporated and the proper treatment of the path integral measure is taken into account one finds that, for particular choices of the DeWitt metric on the space of metrics (in fact, the same choices as in the massless case), one obtains the opposite bound on the graviton mass.

\end{titlepage}

\newpage
\pagestyle{plain}
\setcounter{page}{1}
\newpage

\section{Introduction}

A minimal modification of gravity with some cosmological interest \cite{ah} consists of incorporating a (sufficiently small) graviton mass. There are several problems with this approach, however, the first being the well-known van~Dam-Veltman-Zakharov (vDVZ) discontinuity that persists for arbitrarily small graviton mass \cite{vdvz1, vdvz2}.
It has been shown that it is possible to circumvent this problem in the presence of a (positive or negative) cosmological constant \cite{porr, kmp, hig1}, then, however, another problem of consistency appears: in the case of a positive cosmological constant, in particular for de~Sitter spacetime, one encounters the Higuchi instability for $m^2 < 2 H^2$, where $m$ is the graviton mass and $H$ the Hubble expansion rate \cite{hig1}. It is argued that a scalar degree of freedom propagates then with the wrong sign of the kinetic term, a ghost. The analysis of \cite{hig1} has been done in the canonical quantization formalism; here I consider this problem from the path integral point of view in order to see how it appears in this formalism and investigate whether there are ways to circumvent it.

In pure (massless) Euclidean quantum gravity one encounters a (related, I will argue) problem of stability, known as the conformal factor problem \cite{ghp, mm}, consisting in the unboundedness of the functional integral on the trace part of the perturbation. In \cite{mm} (see also \cite{green}) this problem was analyzed using the generalized DeWitt metric on the space of metrics and with the appropriate treatment of the measure for the path integral, and it was shown that, for particular choices of the metric on the conformal (trace) part, the instability problem disappears.

Here I will apply the same analysis in the case of massive gravity with a cosmological constant (especially for a de~Sitter background). The Bianchi identities together with the equations of motion imply that the graviton field satisfies a constraint, irrespectively of the sources. After a field redefinition that disentangles this constraint one can see that the problem of stability here is also related to the correct definition of the measure for the functional integral. One can see how the problem of the Higuchi instability appears in the path integral formalism and, interestingly, one can check that for the {\itshape same} choices of the DeWitt metric as in the massless case, one obtains the opposite bound for stability of the massive graviton.

It should be stressed, however, that the conformal factor problem of Euclidean (massless) quantum gravity is used mainly as a motivation and the rotation to Euclidean signature is not necessary (in fact, it was not done in \cite{mm} either). There are similarities between the two cases, but since they are two physically different theories (one is massless, gauge-invariant, with two propagating degrees of freedom, and the other massive, without gauge-invariance, and five propagating degrees of freedom) it is still not clear whether the limit of the graviton mass going to zero can be taken consistently. The introduction of the DeWitt metric, with an arbitrary value of its parameters, is also an ad hoc procedure that should be corroborated by some other non-perturbative or canonical method. I believe, however, that the results and methods used here are useful in elucidating the stability problems of massive gravity and offer clues as to the possible ways that they may be circumvented.

In Section~2 I start by describing the path integral treatment of the conformal factor problem in the massless case \cite{mm} and continue with the massive case on the same footing. The incorporation of the path integral measure is critical in obtaining the correct degrees of freedom in both cases and in elucidating the problems of stability for the massive gravitons. The Higuchi instability appears, not as a propagating ghost mode, but as an inverted potential term, and, as mentioned above, it depends on the choices of the DeWitt metric.

In Section 3 I conclude with some comments.

\section{Massive gravitons with a cosmological constant}

The Einstein-Hilbert action with a cosmological constant,
\begin{equation}
S_{EH} = \frac{1}{16\pi G} \, \int_x \,(R - 2\Lambda),
\end{equation}
is linearized around a background metric $g_{\mu\nu}$. The conventions for the signature are (-+++) and $\int_x$ denotes $\int \! d^4 \! x \sqrt{-g}$, with $g=\det{g_{\mu\nu}}$. In terms of the metric perturbation $\delta g_{\mu\nu}= \kappa h_{\mu\nu}$, where
$\kappa^2 =32 \pi G$, and $h=g^{\mu\nu} h_{\mu\nu}$, the quadratic part of the action becomes
\begin{eqnarray}
S_2 &=& \int_x \,\biggl[ \, \frac{1}{2} h^{\mu\nu} \Box h_{\mu\nu} - \nabla^{\mu}h_{\mu\nu} \, \nabla^{\nu} h
      + \nabla^{\mu} h_{\mu\nu} \, \nabla_{\alpha}h^{\alpha\nu} -\frac{1}{2} h \Box h
     \nonumber \\
 &-& \!\!\! h^{\mu\nu} (R_{\alpha\mu\nu\beta} - g_{\mu\beta} R_{\alpha\nu}) h^{\alpha\beta} \!- h R_{\mu\nu} h^{\mu\nu}
 +\!\left(\Lambda - \frac{R}{2}\right)\!(h_{\mu\nu}h^{\mu\nu} - \frac{1}{2} h^2)\biggr].
 \label{S2}
\end{eqnarray}
The proper quantization of the theory is performed when the linear part of the perturbed action vanishes by virtue of the field equations \cite{duff},
\begin{equation}
R_{\mu\nu} = \Lambda \, g_{\mu\nu},
\end{equation}
and the action is gauge invariant with respect to the gauge transformations
\begin{equation}
h_{\mu\nu} \rightarrow  h_{\mu\nu} + \nabla_{\mu} \epsilon_{\nu} + \nabla_{\nu} \epsilon_{\mu}.
\label{gi}
\end{equation}
Massive gravitons are obtained when one adds the Pauli-Fierz mass term,
\begin{equation}
S_{PF} = -\frac{m^2}{2} \int_x \left( h_{\mu\nu}h^{\mu\nu} - h^2 \right),
\label{PF}
\end{equation}
which breaks the gauge invariance (\ref{gi}).

The space of metric perturbations may be analyzed by using the York decomposition \cite{york, mm},
\begin{equation}
h_{\mu\nu} = H_{\mu\nu}+ (L\xi)_{\mu\nu} + \frac{h}{4} g_{\mu\nu},
\label{york}
\end{equation}
where $H_{\mu\nu}$ is transverse and traceless,
\begin{equation}
(L\xi)_{\mu\nu} = \nabla_\mu \xi_\nu + \nabla_\nu \xi_\mu - \frac{1}{2} g_{\mu\nu} \nabla \cdot \xi,
\label{L}
\end{equation}
$h$ is further decomposed as
\begin{equation}
\frac{h}{4} = 2 \sigma + \frac{1}{2} \nabla \cdot \xi,
\label{h}
\end{equation}
and $\xi_\mu$ as
\begin{equation}
\xi_\mu = V_\mu + \nabla_\mu \psi,
\label{xi}
\end{equation}
with $\nabla^\mu V_\mu =0$.

In terms of these variables the quadratic action for the massless case becomes
\begin{equation}
S_2 = \int_x \left[ -\frac{1}{2} H^{\mu\nu} \Delta_2 H_{\mu\nu} +
12 \sigma \left(\Delta_0 -\frac{R}{3}\right) \sigma \right],
\label{S22}
\end{equation}
where $\Delta_0 = - \nabla^2$, $\Delta_2 = \Delta_L - 2\Lambda$ and
\begin{equation}
\Delta_L \phi_{\mu\nu} = -\nabla^2 \phi_{\mu\nu} - 2R_{\mu\rho\nu\sigma}\phi^{\rho\sigma} +R_{\mu\rho}\phi^{\rho}_{\nu} + R_{\nu\rho} \phi^{\rho}_{\mu}
\end{equation}
is the Lichnerowicz operator acting on second rank tensors.

The action is explicitly gauge invariant and one can see that, upon Euclidean continuation, the second term, coming from the trace part, is unbounded, leading to the old conformal factor problem. The story is not complete, however, since one has to take care of the functional measure in the path integral (and its Euclidean continuation)
\begin{equation}
\int [dh_{\mu\nu}] \exp{(i S_2)}.
\end{equation}
It should be stressed also that the rotation to Euclidean signature is not necessary, and the discussion here will continue in Lorentzian signature, one merely needs a way to see how and why the sign of the trace part is flipped.
In \cite{mm} this was done by using the DeWitt metric on the space of metric deformations, which gives a scalar product
\begin{eqnarray}
<\delta g_{\mu\nu}, \delta g_{\alpha\beta}> &=& \int_x G^{\mu\nu,\alpha\beta}\delta g_{\mu\nu}\delta g_{\alpha\beta}
\nonumber \\
&=& \int_x  \left[ H^{\mu\nu}H_{\mu\nu} + \xi^{\alpha}(\Delta_1\xi)_{\alpha} + \frac{(1+2C)}{4} h^2 \right],
\label{dewitt}
\end{eqnarray}
where
\begin{equation}
G^{\mu\nu,\alpha\beta} = \frac{1}{2} \left( g^{\mu\alpha}g^{\nu\beta} + g^{\nu\alpha}g^{\mu\beta} + C g^{\mu\nu} g^{\alpha\beta} \right)
\label{dw}
\end{equation}
is the general DeWitt metric that depends on an arbitrary constant $C$ and
\begin{equation}
(\Delta_1 \xi)_\alpha = -2\nabla^2 \xi_\alpha - \nabla_\alpha \nabla^\beta \xi_\beta -2 R_\alpha^\beta \xi_\beta.
\end{equation}
The defining condition
\begin{equation}
\int [d\delta g] \exp{(-\frac{i}{2} < \delta g, \delta g >)} = 1
\end{equation}
gives the transformation of the measure as
\begin{equation}
[dh_{\mu\nu}] = [dH_{\mu\nu}] ({\det}^\prime \Delta_1)^{1/2} [d\xi_\mu][d h]
\end{equation}
and, since
\begin{equation}
(\Delta_1 V)_\alpha = 2 \left( \Delta_0 - \frac{R}{4} \right) V_\alpha,
\end{equation}
\begin{equation}
(\Delta_1 \nabla\psi)_\alpha = 3 \nabla_\alpha \left( \Delta_0 -\frac{R}{3}\right) \psi,
\end{equation}
as \cite{mm}
\begin{equation}
[dh_{\mu\nu}] = [dH_{\mu\nu}] \left({\det}^\prime \left( \Delta_0 - \frac{R}{4} \right)\right)_T^{1/2} [dV_\mu]
\left({\det}^\prime \left( \Delta_0 -\frac{R}{3}\right)\right)_S^{1/2}[d\sigma][d\psi].
\label{meas1}
\end{equation}
The primes on the determinants denote that we omit any zero or negative eigenvalues upon continuation to Euclidean space, the first determinant is on the space of transverse vectors and the second on the scalars.

The transverse and traceless tensor $H_{\mu\nu}$ has five degrees of freedom, three of which are removed by the vector determinant, leaving the two degrees of freedom of a massless graviton. The remaining scalar, $\sigma$, in (\ref{S22}) does not correspond to a propagating degree of freedom: after the (non-local) redefinition, $\chi =({\det}( \Delta_0 -\frac{R}{3}))^{1/2}\,\sigma$, in order to incorporate the scalar determinant, the apparently kinetic term for $\sigma$ becomes a potential term for $\chi$. The only remnant of the instability is the fact that this potential term is inverted (unbounded below) when $C>-\frac{1}{2}$, but becomes of the usual sign and bounded when $C<-\frac{1}{2}$ in (\ref{dewitt}), corresponding to an opposite continuation of the conformal factor. More details can be found in \cite{mm}, here I will continue to apply this procedure in the massive case, that is to consider the combined action
$S = S_2 + S_{PF}$.

When the Pauli-Fierz term is added to the Einstein-Hilbert action it has the effect of propagating a massive graviton with five degrees of freedom and the full action is not gauge invariant. In terms of the variables used here one has
\begin{equation}
S_{PF}=-\frac{m^2}{2}\!\int_x \left[ H^{\mu\nu}H_{\mu\nu} + 2V^\mu\!\!\left(\Delta_0 - \frac{R}{4}\right)\!V_\mu
      + 4 \Lambda \psi \nabla^2\psi -24 \sigma\nabla^2\psi - 48 \sigma^2 \right].
\label{pf2}
\end{equation}
The $\psi\nabla^2\psi$ term here also has an opposite (ghostlike) sign, like the $\sigma\nabla^2\sigma$ term in (\ref{S22}), and there is also a mixing term that has to be considered in order to investigate the stability of the theory. Before doing so, however, we note that the Bianchi identities, together with the equations of motion for a massive graviton, imply the relation
\begin{equation}
\nabla^\mu \nabla^\nu h_{\mu\nu} = \nabla^2 \, h,
\label{eq1}
\end{equation}
or
\begin{equation}
3\nabla^2\sigma = \Lambda\nabla^2 \psi,
\label{eq2}
\end{equation}
which holds irrespectively of the sources. This may be used as a constraint in order to disentangle the dynamics of the theory and one may define the field
\begin{equation}
\Phi = 3\sigma - \Lambda\psi,
\label{eq3}
\end{equation}
which is non-dynamical, constrained by $\nabla^2\Phi =0$. After collecting the various terms from (\ref{S22}) and (\ref{pf2}) we get for the full action:
\begin{eqnarray}
S=\int_x\biggl[&-&\frac{1}{2} H^{\mu\nu} (\Delta_2+m^2) H_{\mu\nu} -m^2V^\mu\left(\Delta_0 - \frac{R}{4}\right)V_\mu
     \nonumber\\
        &-&18\left(\frac{m^2}{\Lambda}-\frac{2}{3}\right)\sigma\left(\Delta_0 -\frac{R}{3}\right)\sigma\biggr]
\label{final}
\end{eqnarray}
plus terms of the form $\sigma\nabla^2 \Phi$, $\Phi\nabla^2 \Phi$ and related mass terms involving $\Phi$. Since $\Phi$ is constrained these additional terms are irrelevant to the dynamics of the theory.

One sees in (\ref{final}) that, for $m^2 < 2 H^2$ ($\Lambda = 3 H^2$ and $R=4\Lambda$ as usual), the $\sigma$ field seems to acquire a wrong sign, ghostlike, kinetic term, signaling the Higuchi instability. Again, the story is not complete, however, since one has to take care of the path integral measure. In fact, the incorporation of the vector determinant from (\ref{meas1}) is vital in order to cancel the vector term in (\ref{final}), leaving the five degrees of freedom of $H_{\mu\nu}$, as is expected for a massive graviton. The (non-local) redefinition, $\chi_\mu =({\det}(\Delta_0-\frac{R}{4}))^{1/2}\,V_\mu$, transforms the vector kinetic term into a potential term (with the correct sign, bounded).

The change of variables in (\ref{eq3}) has trivial Jacobian, so the functional integral over $[d\sigma][d\psi]$ in (\ref{meas1}) can be done over $[d\sigma][d\Phi]$ and another redefinition, $\chi =({\det}(\Delta_0-\frac{R}{3}))^{1/2}\sigma$, transforms the last term in (\ref{final}) into a potential term, which is, however, unbounded below for $m^2 < 2 H^2$. Thus, the Higuchi instability in the path integral formalism does not correspond to a propagating ghost but, rather, an inverted potential term. It should be clear that, if the same treatment as the massless case is applied here, that is, if one considers a DeWitt metric with $C < -\frac{1}{2}$, then this has the effect of reversing the previous conclusion and massive gravitons become stable for $m^2 < 2 H^2$, a result which is more interesting from the cosmological point of view.

\section{Comments}

The consistency of massive gravity in general curved backgrounds is an interesting problem because of various possible cosmological and theoretical considerations \cite{ah2, def1, def2, porr, kmp, duff2, d1, izumi, koyama}. Various works deal with this problem, mostly from the Hamiltonian (canonical) point of view \cite{hig1, deser, grisa, blas, gab2}. Here I considered the path integral approach, following related work \cite{mm} in the massless case, and showed how the Higuchi instability appears in this formalism. The appropriate treatment of the path integral measure is critical in this approach and it is interesting that, if a particular definition of the measure in terms of the DeWitt metric is employed, similarly to the massless case, one gets the opposite bound for the graviton mass.

Naturally it would be interesting to reconcile the two approaches, hopefully to derive the path integral measure with a specific value of the DeWitt metric (which remains arbitrary in this respect) from the more fundamental, Hamiltonian, point of view. In any case, however, the path integral approach gives some additional insight to the problems considered.

\vspace{0.2in}

\centerline{\bf Acknowledgements}
\noindent
This work was completed while visiting the National Technical
University of Athens. I would like to thank the people of the Physics
Department for their hospitality.

\end{document}